\documentclass[twocolumn,amssymb,fleqn]{revtex4} 
\usepackage{epsfig,amssymb,amsmath,graphicx,subfigure,hyperref  }  
\usepackage{mathrsfs}
\usepackage{color}
\usepackage{multirow}
\usepackage{tabularx}

\begin{document}

\title{Reflection Phase Shift of One-dimensional Plasmon Polaritons in Carbon Nanotubes} 
\author{Xingdong Luo,$^{1,2,\dag}$ Cheng Hu,$^{1,2,\dag}$ Bosai Lyu,$^{1,2,\dag}$ Liu Yang,$^{4,\dag}$ Xianliang Zhou,$^{1,2}$ Aolin Deng,$^{1,2}$ Ji-Hun Kang$^{3,*}$ and Zhiwen Shi$^{1,2,}$} 
\email{jihunkang@kongju.ac.kr; zwshi@sjtu.edu.cn}
\affiliation{
$^1$Key Laboratory of Artificial Structures and Quantum Control (Ministry of Education), Shenyang National Laboratory for Materials Science, School of Physics and Astronomy, Shanghai Jiao Tong University, Shanghai 200240, China\\
$^2$Collaborative Innovation Center of Advanced Microstructures, Nanjing 210093, China\\
$^3$Department of Optical Engineering, Kongju National University, Cheonan 31080, Korea\\
$^4$ School of Physics and Astronomy, University of Manchester, Manchester M13 9PL, United Kingdom} 
\begin{abstract}
We investigated, both experimentally and theoretically, the reflection phase shift (RPS) of one-dimensional plasmon polaritons. We launched 1D plasmon polaritons in carbon nanotube and probed the plasmon interference pattern using scanning near-field optical microscopy (SNOM) technique, through which a non-zero phase shift was observed. We further developed a theory to understand the nonzero phase shift of 1D polaritons, and found that the RPS can be understood by considering the evanescent field beyond the nanotube end. Interesting, our theory shows a strong dependence of RPS on polaritons wavelength and nanotube diameter, which is in stark contrast to 2D plasmon polaritons in graphene where the RPS is a constant. In short wave region, the RPS of 1D polaritons only depends on a dimensionless variable -- the ratio between polaritons wavelength and nanotube diameter. These results provide fundamental insights into the reflection of polaritons in 1D system, and could facilitate the design of ultrasmall 1D polaritonic devices, such as resonators, interferometers.
\end{abstract}

\maketitle
\section{Introduction}
{\huge\textbf{R}}eflection phase shift (RPS) of propagating electromagnetic modes at interfaces is a fundamental physical concept. The reflection phenomena of the traditional plane wave propagating in three-dimensional dielectric environment are well described by Fresnel reflection and transmission coefficients\cite{novotny2012principles,jackson1999classical}. However, RPS of polaritons, which are propagating electromagnetic modes confined in low-dimensional materials, has not received much attention for a long time. With the emergence of low-dimensional materials in recent years, such as graphene, hexagonal boron nitride and carbon nanotube, it has been possible to achieve high quality polaritons in experiments\cite{low2017polaritons,basov2016polaritons,ni2018fundamental,dai2014tunable,fei2012gate,woessner2015highly,ni2016ultrafast,chen2012optical,ju2011graphene,ni2015plasmons,giles2018ultralow,shi2015observation,hu2017imaging}, which provides a golden opportunity to study RPS of polaritons in low-dimensional systems. Until now RPS of two-dimensional (2D) polaritons has already been studied\cite{prbTony2014,nanolettKang2017}. It has been reported that RPS of plasmon polariton in graphene is near 0.25$\pi$\cite{nanolettKang2017}, which seems anomalous because usually a simple 0 or $\pi$ phase shift is expected when reflection occurs at edges where there is a sudden disappearance of surface conductivity. This strange RPS can be explained by the Goos-hanchen phase shift and is related to electromagnetic energy stored in evanescent field\cite{prbTony2014,nanolettKang2017}. RPS of one-dimensional polaritons in ultra-thin 1-D system, however, has rarely been studied yet. Here, we combine infrared nanoimaging and numerical simulation to achieve the RPS of 1-D plasmon polaritons in metallic carbon nanotube, where a non-zero RPS is observed. We further developed a theoretical model to understand the observed nonzero phase shift and found that the RPS is also related to the electromagnetic energy stored in the near-field beyond the nanotube end, similar to that of 2D plasmons in graphene. Our theory shows that the RPS of 1D polaritons has a strong dependence on the plasmon wavelength, as well as the diameter of carbon nanotube, which is distinct from its 2-D counterparts.

Plasmon polaritons confined in metallic carbon nanotube exhibit extraordinary quality. The 1-D Dirac electron in metallic carbon nanotube preserves its chiral handedness when moving forward\cite{ando1998berry}. Such chiral electrons are not susceptible to the slowly varied charge impurity potential or acoustic phonons\cite{mceuen1999disorder,suzuura2002phonons}, leading to the elimination of all loss channels in one-dimensional carbon nanotube. Besides, due to the strong correlated state of the electron system, the 1-D plasmons behave as luttinger-liquid plasmons qualitatively differing from classical plasmon excitations\cite{shi2015observation,wang2019logarithm}. The phase velocities of Luttinger-liquid plasmons are stable, independent of carrier concentration or excitation wavelength. Due to the Luttinger-liquid plasmons in carbon nanotube can simultaneously achieve extraordinary spatial confinement and high-quality factor, the metallic carbon nanotube has become an excellent candidate for studying RPS of 1-D plasmon polaritons.
\begin{figure}
\centering
\includegraphics[width=0.35\textwidth]{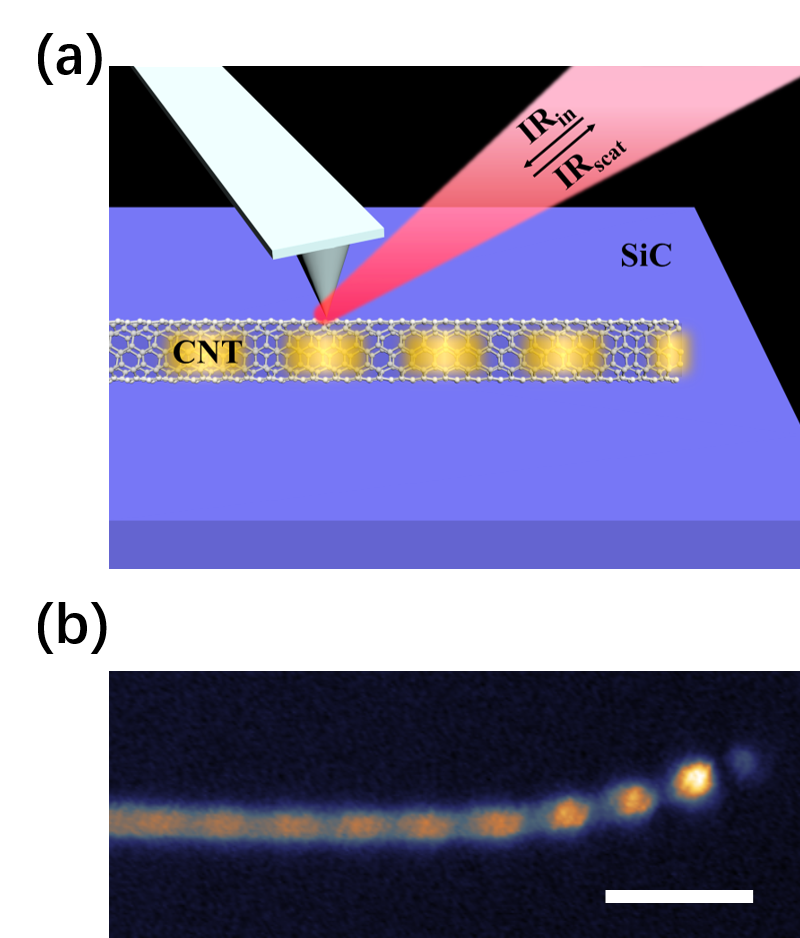}
\caption{\textbf{One-dimensional plasmon polaritons in carbon nanotube} (a) Generating 1-D plasmon polaritons with s-SNOM. (b) Near-field image of launched-reflected plasmons interference. The incident wavelength is $10.6\mu m$. Scale bars, 150 nm.}
\end{figure}
\section{Experiment}
Experimentally we utilized infrared scattering-type scanning near-field optical microscopy (s-SNOM) to launch 1-D plasmon polaritons in carbon nanotube and to probe detailed plasmon interference patterns, as shown in FIG.1a. Single-wavelength infrared light in the range of $7.7\mu m\sim 9.6\mu m$ was focused onto the apex of a gold-coated atomic force microscope (AFM) tip. The sharp AFM tip enabled optical excitation of plasmons due to its large near-field momentum. The excited plasmon wave propagated along the nanotube and was reflected by the nanotube end. The interference between the tip-launched plasmon and the reflected plasmon produced a periodic electromagnetic fields distribution, which modified the intensity of the tip-scattered infrared radiation measured by an HgCdTe detector in the far field. A schematic of near-field excitation and probing is displayed in FIG.1a. Through probing the near-field interference pattern, we can achieve not only the plasmon wavelength but also the phase shift of plasmon reflection. An experimental near-field nano-imaging of plasmons in a carbon nanotube is shown in FIG.1b.

We systematically investigated  RPS of 1-D plasmon polaritons in metallic carbon nanotubes, with the wavelength of incident infrared light ranging from $7.7\mu m\sim9.6\mu m$. FIG.2a shows the near-field image as well as the corresponding AFM topography image, from which the diameter of the nanotube was identified to be 2nm. Also, the AFM topography image allows for location of nanotube end, labeled by a red bar. The oscillating infrared scattering intensity along the nanotube in FIG.2b reveals how the plasmon polaritons get reflected and damped as it propagates. For the sake of an accurate determination of plasmon wavelength $\lambda_p$ and reflection phase shift $\phi$, an oscillator form $e^{-2\pi x/(Q\lambda_p)}\cos{(4\pi x/\lambda_p+\phi)}$ was used to fit the experimental curve, from which we obtain $\lambda_p=91nm$ and $\phi=0.26\pi$ for excitation wavelength of $7.8\mu m$. Please note that the plasmon in CNT exhibit  an extraordinary spatial confinement ($\lambda_p/\lambda_0\sim1/100$) and a relatively high quality factor ($Q > 10$), which makes it a promising candidate for future nanoscale photonic devices. FIG.2c shows the value of RPS extracted from FIG.2b. (See Supplementary Material for details). We found that an anomalous RPS around $0.2\pi$ appeared in this one-dimensional system, which deserves further theoretical investigation for its origin. In the following, we showed that this RPS is linked to the evanescent field beyond carbon nanotube end, with a quantitively description of the relation between energy flow of the evanescent field and the corresponding RPS.
\begin{figure*}
\centering
\includegraphics[width=1.0\textwidth]{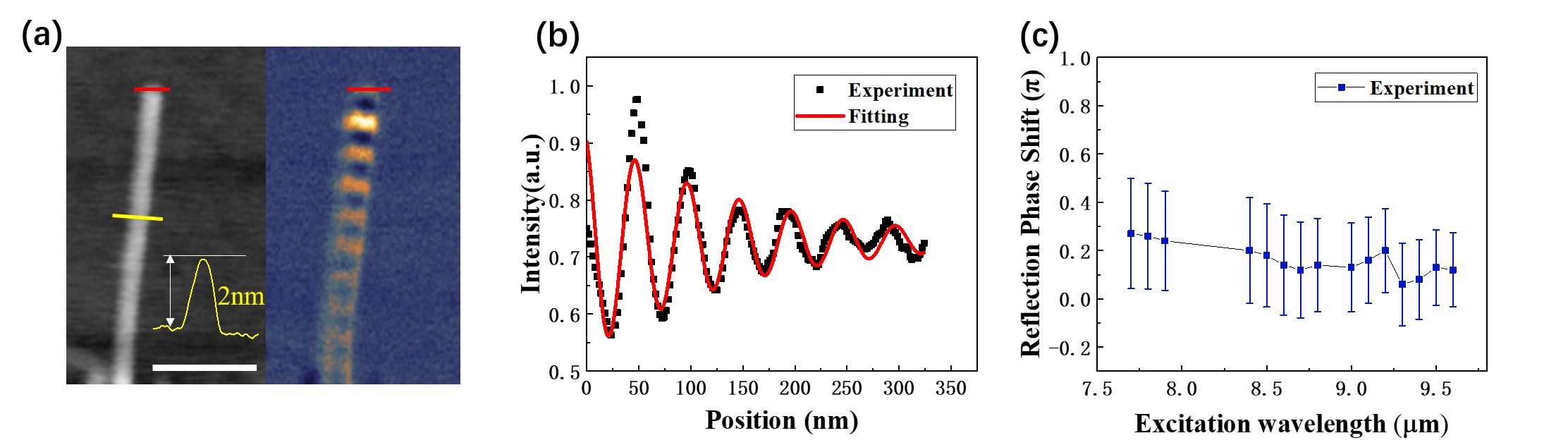}
\caption{\textbf{Reflection phase shift (RPS) of edge-reflected 1-D plasmon polaritons.} (a) AFM topography image (Left) and near-field image (Right) of a representative nanotube sample. The incident wavelength is $7.8\mu m$. Scale bars, 150 nm. Inset is the line profile across nanotube along the yellow line. From the height profiles, diameter of the nanotube was determined to be 2nm. (b) Near-field response line profile taken along the nanotube in the near-field image. (c) RPS of 1-D plasmons with the wavelength of incident infrared light ranging from $7.7\mu m\sim 9.6\mu m$.}
\end{figure*}
\section{Theory and Discussion}
We investigated RPS of one-dimensional plasmon polaritons theoretically using a combination of analytical calculations and COMSOL simulations. We first recreated the reflection phenomena of 1-D polaritons in carbon nanotube system by solving Maxwell’s equation numerically with COMSOL. Considering the fact that there was no accurate optical conductivity of carbon nanotube reported in previous literatures, we adjusted the optical conductivity of the nanotube to make the simulated plasmon wavelength match with the experimental value for each incident wavelength.(Shown in FIG.3a) This is reasonable because usually there is a one to one correspondence between polaritons wavelength and imaginary part of optical conductivity. And we simultaneously tuned the real part of optical conductivity to obtain long range well-defined interference patterns. In FIG.3a, we display simulated plasmon interference pattern near the end of a metallic carbon nanotube with diameter of $2nm$ for excitation wavelength at $6.1\mu m$, $7.1\mu m$, $8.2\mu m$, $9.1\mu m$ and $10\mu m$ from top to bottom panels. Those interference patterns clearly show that the first node is not located at $\lambda_p /4$ positon, indicating a non-zero RPS. The RPS extracted from those simulation pattern is ploted in FIG.3c, which qualitatively agrees with the experimental data.\\
\begin{figure*}
\centering
\includegraphics[width=1.0\textwidth]{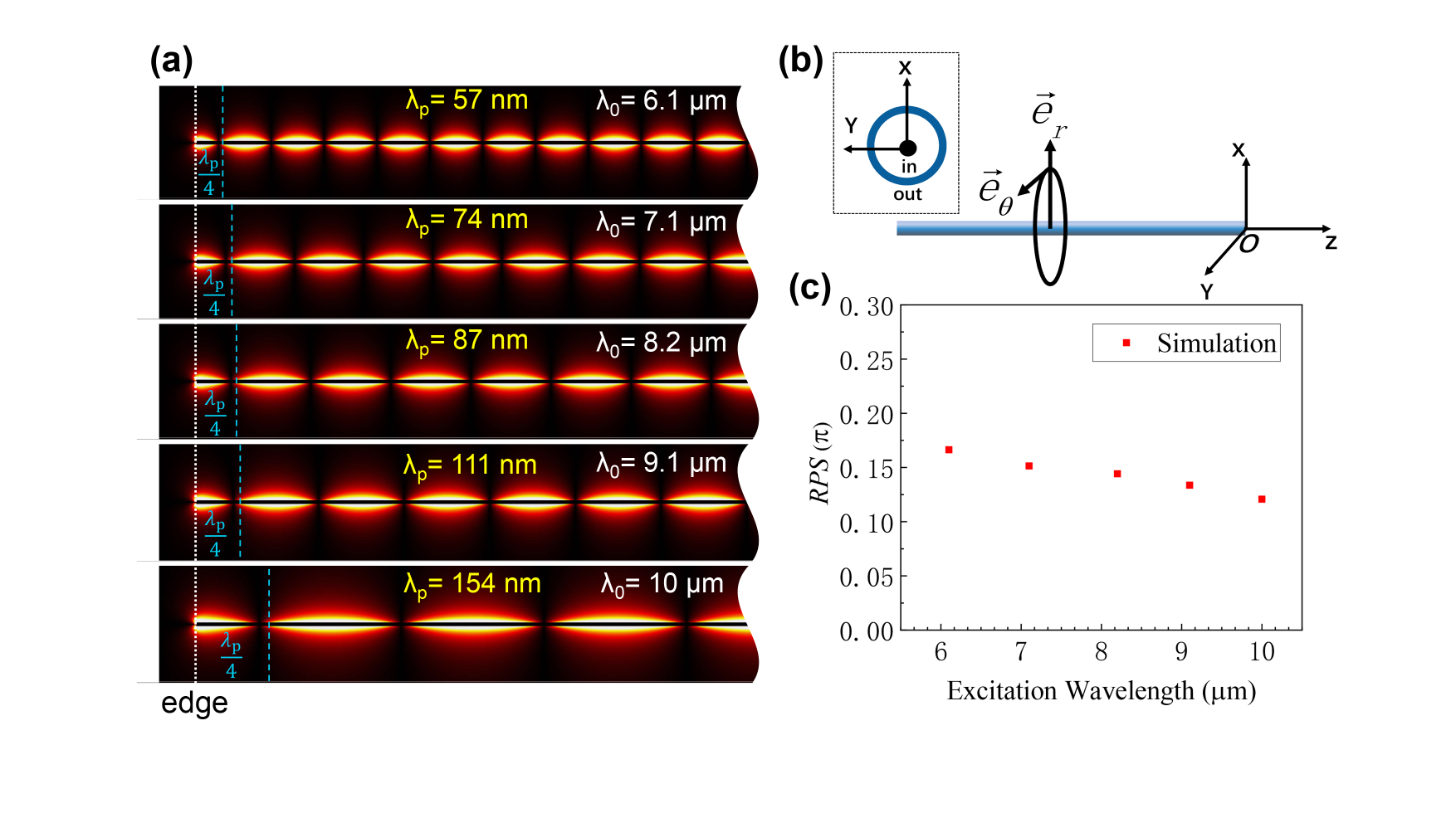}
\caption{\textbf{Simulation of 1-D plasmon polaritons in carbon nanotube system.} (a) Interference patterns of the $r-$component of electric fields of launched-reflected 1-D plasmon polaritons in carbon nanotube. (b) Schematic of the modelled system. (c) RPS of 1-D polaritons derived from (a).}
\end{figure*}

To get a deep insight of the anomalous RPS and the corresponding evanescent fields in the system, an analytical calculation was proposed by modeling the reflection problem in an extreme thin cylindrical shell, as shown in FIG.3b. The reflection problem can be dealt with as a boundary problem of the electromagnetic fields at $z=0$ plane. The launched plasmon polaritons mode manifests itself as modified Bessel function according to both Maxwell’s equations and the rotational symmetry of the system.

\begin{align}
&\mathbf{H}^L(\mathbf{r})=h(r)e^{ik_pz}\mathbf{e}_\theta\label{eq1}\\
&h(r)= \left\{ \begin{array}{ll}
C I_1(K_{pr}r),&r<r_0\\\\
K_1(K_{pr}r),&r>r_0
  \end{array} \right.\nonumber\\
  &K_{pr}=\sqrt{k_p^2-k_0^2}\nonumber
\end{align}

where $\mathbf{H}^L(\bold{r})$ is the magnetic field of launched polaritons mode, which rotates around the carbon nanotube, $k_p$ is polaritons wavevector, and $k_0$ is excitation wavevector in vacuum. For strongly confined 1-D polaritons in our system, $k_p$ is much larger than $k_0$. $I_1(K_{pr}r)$ is modified Bessel function of the first kind, and $K_1(K_{pr}r)$ is modified Bessel function of the second kind. $r_0$ is the radius of carbon nanotube. $C$ is an undetermined coefficient, which should be determined by the continuity of z-component of the electric field at $r=r_0$. Using the Maxwell equations, the corresponding electric field reads
\begin{align}
\mathbf{E}^L&=\frac{i}{\omega\varepsilon_0}\nabla\times\mathbf{H}^L\nonumber\\
&=\frac{i}{\omega\varepsilon_0}[K_{pr}g(r)\mathbf{e}_z-ik_ph(r)\mathbf{e}_r]e^{ik_pz}\label{2}\\
&g(r)= \left\{ \begin{array}{ll}
C I_0(K_{pr}r),&r<r_0\\\\
-K_0(K_{pr}r),&r>r_0
  \end{array} \right.\nonumber
\end{align}

Considering that only the r-component of electric field was detected in our experiment, we would seek for the launched-reflected polaritons interference in the following form 
\begin{align}
E^p_r(r,z)&=\frac{k_p}{\omega\varepsilon_0}h(r)(e^{ik_pz}+Re^{-ik_pz})\label{3},\\ 
H^p_\theta(r,z)&=h(r)(e^{ik_pz}-Re^{-ik_pz}).\label{4}
\end{align}
Where $E_r^p(r,z)$ is the $r-$component of the electric field of polaritons, $H^p_\theta(r,z)$ is the $\theta-$component of the magnetic field of polaritons, and $R$ is reflection coefficient. RPS of 1-D polaritons is defined as $\arg(R)$. For convenience, the polaritons modes are denoted to be the electromagnetic fields in the region $z<0$\cite{prbTony2014}, 
\begin{align}
  E^<_r(r,z)&\approx E^p_r(r,z) \\
     H^<_\theta(r,z)&\approx H^p_\theta(r,z).
  \end{align}
We note that the reflection of the 1D polaritons induces evanescent waves near the end of the carbon nanotube. In the carbon nanotube region, those waves can be expanded in terms of an electromagnetic continuum of unbounded modes, as those of the 2D polaritons are introduced in refs.\cite{nanolettKang2017} and ref.\cite{chaves2018scattering}. In the 2D case, the unbounded modes modify the phase shift from $0.3\pi$ to $0.25\pi$ \cite{nanolettKang2017}, and similar improvement is expected in our 1D case by employing the unbounded modes.

Now let us represent the field in the region $z > 0$ in the form of the Fourier-Bessel expansion
\begin{align}
    &H^>_\theta(r,z)=\int_0^{+\infty}kdke^{ik_zz}J_1(kr)f(k),\\
   & E^>_r(r,z)=\frac{1}{\omega \varepsilon_0}\int_0^{+\infty}k_zkdke^{ik_zz}J_1(kr)f(k),\\
    &k_z=\sqrt{k_0^2-k^2}.\nonumber
  \end{align}
  With the aid of boundary conditions in the plane $z=0$, we obtain 
  \begin{align}
&\frac{1-R}{1+R}=\frac{r_0^2}{\int_0^{+\infty}rdrh^2(r)}\int_0^{+\infty}kdk\frac{G^2(k)}{k_zk_p},\label{9}
\end{align}
where $G(k)$ is a dimensionless function, the concrete expression of which can be found in  Supplementary Material.\\

   Integrals in Eq.(\ref{9}) can be computed numerically. FIG.4a shows the dependence of the calculated RPS on polariton wavelengths for carbon nanotubes with different diameters, and FIG.4b displays the corresponding reflection amplitudes. These results agree well with the former discussion. The reflection amplitudes are nearly equal to unity, in accordance with the fact that the density of states on the polaritons channel is much larger than the density of photonic radiation modes\cite{shi2015observation,moradi2007collective}.
The slight difference between analytical calculation and simulation in FIG.4c can be attributed to the effect of the unbounded modes, which reflect the complex behavior of electromagnetic fields close to the nanotube end\cite{lalanne2006interaction,nikitin2009diffraction}.\\

For the sake of a better description of RPS of 1-D quantum plasmon polaritons in carbon nanotubes, we further investigate our theory in the short-wave region $(k_p\gg k_0)$, which is also the region for most quantum plasmon polaritons in carbon nanotubes\cite{shi2015observation}. When $k_p\gg k_0$, Eq.(\ref{9}) was reduced to (See Supplementary Material)
\begin{align}
   &\frac{1-R}{1+R}=F(\frac{\lambda_p}{D}) 
  \end{align}
$D$ is the diameter of the nanotube. We can see in the short wave region, the reflection coefficient only depends on the ratio between polaritons wavelength and nanotube diameter.

For polaritons in 2-D materials, the reflection phase shift is almost independent with the thickness of materials as long as the thickness is much smaller than polariton wavelength\cite{prbTony2014,nanolettKang2017}. The RPS of 2-D polaritons is nearly $0.25\pi$ in the short-wave region. However, here even in short-wave region the RPS of 1-D polaritons has a strong dependence on the ratio between polaritons wavelength and diameter of the nanotube, as shown in FIG.4d. This is likely due to the difference in geometry of 1-D and 2-D polaritons modes.\\
\begin{figure*}
\centering
\includegraphics[width=0.8\textwidth]{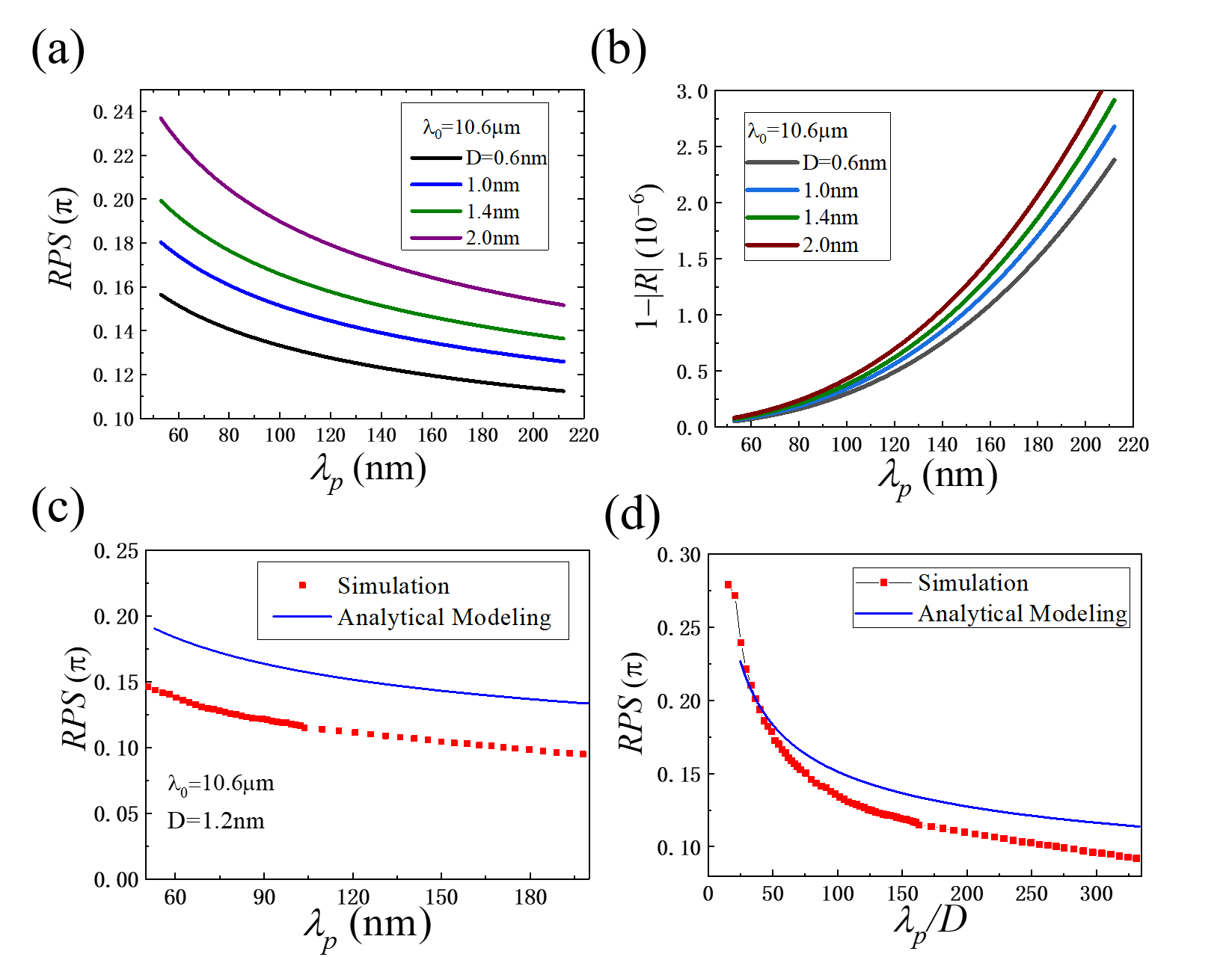}
\caption{\textbf{Theoretical results of 1-D polaritons reflection.} (a) Analytically calculated RPS for various polaritons wavelengths and diameters of nanotube. The wavelength of incident infrared light is 10.6um. (b) Plot of the  corresponding reflection amplitudes $\left|R\right|$. (c) The comparation between analytical and simulation results. (d) The results for short polaritons wavelength. In short-wave approximation, the RPS only depends on the relative size between polaritons wavelength and diameter of the nanotube.}
\end{figure*}

It is desirable to investigate the evanescent field in the nanotube system. Previous study of RPS of 2-D polaritons in graphene has shown that the phase shift is related to the evanescent field according to the Goos-Hänchen theory. We believe that anomalous RPS in 1-D system is also related to the evanescent field beyond nanotube end within a few nanometers. A notable portion of electromagnetic energy can be temporally stored in the evanescent fields and then get back reflected to be injected into the polaritons. As a result, the reflected polaritons experience an extra phase shift relative to the launched polaritons\cite{jackson1999classical,nanolettKang2017}. To quantitively characterize the energy transport, we calculated the energy flow of the electromagnetic field in the region $z>0$ and predictably found most of the energy cannot be radiated out. (See Supplementary Material for details) So in this system,the RPS can be associated with the energy transport process of the evanescent field.
\begin{align}
   P_{evane}=\bar{P}_0[\cos(2\omega t)-\cos(2\omega t-2\phi)]\label{11}
 \end{align}
$P_{evane}$ is the energy flow of evanescent field at $z=0$ plane and $\bar{P}_0$ is time-average energy flow of launched polaritons at $z=0$ plane. Eq.(\ref{11}) reveals the connection between RPS ($\phi$) and the evanescent field.\\
\section{Conclusion}
In summary, we have studied the reflection phase shift of 1-D polaritons. Experimentally, we have measured the RPS of 1-D plasmon polaritons in metallic carbon nanotube using s-SNOM technique. Theoretically, we have developed an analytical theory to describe the reflection process of 1-D polaritons. The calculated RPS is consistent with the results from experiment and simulation. We showed that RPS of 1-D polaritons depends on both polaritons wavelength and nanotube diameter. In short wave region for quantum plasmon polaritons in carbon nanotubes, the RPS of 1-D polaritons only depends on a dimensionless variable, the ratio between polaritons wavelength and diameter of the nanotube. Moreover, we analyzed the energy transport process in the system, and associated RPS with the energy flow of the evanescent field. Similar to 2-D plasmon polaritons in graphene, the RPS in 1-D electrons system also originates from existence of the evanescent electromagnetic field at the nanotube end. These results are essential for a variety of designs and applications of plasmonic resonances, such as novel ultrasensitive sensors, heat concentrating agents in cancer treatment, building blocks of metamaterials, and active elements in all-optical signal processing.
\section{Acknowledgement}
This work is supported by National Natural Science Foundation of China (11574204 and 11774224), and National Key Research and Development Program of China (2016YFA0302001). Zhiwen Shi acknowledges support from the National Thousand Youth Talents Plan, the Shanghai Thousand Talents Plan, the Program for Professor of Special Appointment (Eastern Scholar) at Shanghai Institutions of Higher Learning and additional support from a Shanghai talent program. Ji-Hun Kang was supported by the Basic Science Research Program through the National Research Foundation of Korea (NRF) grant funded by the Korea government (MSIP) (NRF-2018R1C1B6009007). Liu Yang acknowledges support from China Scholarship Council Grant No. 201906230305.

\bibliography{phase_paper.bib}
\end{document}